\documentclass[aps,prl,superscriptaddress,showpacs,twocolumn,floatfix]{revtex4-1}
\usepackage{latexsym}
\usepackage{amsthm}
\usepackage{amsmath}
\usepackage{amssymb}
\usepackage{hyperref}
\usepackage{bbm}
\usepackage{color}
\usepackage{colordvi}
\usepackage{comment}
\usepackage{times,latexsym,graphicx,wrapfig,url}
\usepackage{epsfig,lineno,bm}
\usepackage{subfigure}
\usepackage{mciteplus}

\newcommand{\gsim}{\lower.7ex\hbox{$\;\stackrel{\textstyle>}{\sim}\;$}}
\newcommand{\lsim}{\lower.7ex\hbox{$\;\stackrel{\textstyle<}{\sim}\;$}}

\newcommand{\degr}{\ensuremath ^\circ}

\def\epm{e^+e^-}

\newcommand{\MeV}{\,\mathrm{MeV}}

\newcommand{\be}{\begin{equation}}
\newcommand{\ee}{\end{equation}}
\newcommand{\bea}{\begin{eqnarray}}
\newcommand{\eea}{\end{eqnarray}}

\newcommand{\bef}{\begin{figure}[htbp]\begin{center}}
\newcommand{\eef}{\end{center}\end{figure}}

\newcommand{\mc}{\mathcal}
\newcommand\UVa{University of Virginia, Charlottesville, Virginia 22903}
\newcommand\Yerevan{Yerevan Physics Institute, Yerevan 375036, Armenia}
\newcommand\TelAviv{Tel Aviv University, Tel Aviv, 69978 Israel}
\newcommand\FIU{Florida International University, Miami, Florida 33199}
\newcommand\JLab {Thomas Jefferson National Accelerator Facility, Newport News, Virginia 23606}

\newcommand\CalSt{California State University at Los Angeles, Los Angeles, California 90032}
\newcommand\MIT{Massachusetts Institute of Technology, Cambridge, Massachusetts 02139}

\newcommand\UNH{University of New Hampshire, Durham, New Hampshire 03824}
\newcommand\WaM{College of William and Mary, Williamsburg, Virginia 23187}
\newcommand\UMASS{University of Massachusetts, Amherst, Massachusetts 01003}
\newcommand\KU{University of Kentucky, Lexington, Kentucky 40506}

\newcommand\SLAC{SLAC National Accelerator Laboratory, Menlo Park, California 94025}
\newcommand\Long{Longwood University, Farmville, Virginia 23909}
\newcommand\SCU{University of South Carolina, Columbia, South Carolina 29225} 
\newcommand\Perimeter{Perimeter Institute for Theoretical Physics, Waterloo, ON N2L 2Y5, Canada}
\newcommand\GWU{George Washington University, Washington DC 20052}
\newcommand\Kent{Kent State University, Kent, Ohio 44242}
\newcommand\SMU{Saint Mary's University, Halifax, NS B3H 3C3, Canada}
\newcommand\NYU{New York University, New York, New York 10012}
\newcommand\Syra{Syracuse University, Syracuse, New York 13244}
\newcommand\CMU{Carnegie Mellon University, Pittsburgh, Pennsylvania 15213}
\newcommand\NSU{Norfolk State University, Norfolk, Virginia 23504}
\newcommand\Stanford{Stanford University, Menlo Park, California 94025}
\newcommand\NCAandT{North Carolina Agricultural and Technical State University, Greensboro, North Carolina 27411}

\begin{document}
\begin{flushright}{JLAB-PHY-11-1406 / SLAC-PUB-14491}
\end{flushright}
\title{
\vskip -0.5cm
Search for a new gauge boson in the $A'$ Experiment (APEX)}
\author{S.~Abrahamyan}          \affiliation{\Yerevan} 
\author{Z.~Ahmed}               \affiliation{\Syra}
\author{K.~Allada}              \affiliation{\KU} 
\author{D.~Anez}                \affiliation{\SMU} 
\author{T.~Averett}             \affiliation{\WaM} 
\author{A.~Barbieri}	        \affiliation{\UVa} 
\author{K.~Bartlett}            \affiliation{\UNH}
\author{J.~Beacham}             \affiliation{\NYU}
\author{J.~Bono}                \affiliation{\FIU}
\author{J.R.~Boyce}               \affiliation{\JLab} 
\author{P.~Brindza}             \affiliation{\JLab}
\author{A.~Camsonne}            \affiliation{\JLab} 
\author{K.~Cranmer}             \affiliation{\NYU}           
\author{M.M.~Dalton}              \affiliation{\UVa}
\author{C.W.~de~Jager}          \affiliation{\JLab}  \affiliation{\UVa} 
\author{J.~Donaghy}             \affiliation{\UNH}
\author{R.~Essig}               \thanks{rouven@slac.stanford.edu} \affiliation{\SLAC}
\author{C.~Field}               \affiliation{\SLAC}
\author{E.~Folts}               \affiliation{\JLab}
\author{A.~Gasparian}           \affiliation{\NCAandT} 
\author{N.~Goeckner-Wald}       \affiliation{\CMU}
\author{J.~Gomez}               \affiliation{\JLab}
\author{M.~Graham}              \affiliation{\SLAC}
\author{J.-O.~Hansen}            \affiliation{\JLab} 
\author{D.W.~Higinbotham}       \affiliation{\JLab}
\author{T.~Holmstrom}	        \affiliation{\Long} 
\author{J.~Huang}               \affiliation{\MIT} 
\author{S.~Iqbal}               \affiliation{\CalSt} 
\author{J.~Jaros}               \affiliation{\SLAC}
\author{E.~Jensen}              \affiliation{\WaM}
\author{A.~Kelleher}            \affiliation{\MIT} 
\author{M.~Khandaker}           \affiliation{\NSU}      \affiliation{\JLab}   
\author{J.J.~LeRose}              \affiliation{\JLab}
\author{R.~Lindgren}            \affiliation{\UVa}
\author{N.~Liyanage}            \affiliation{\UVa}
\author{E.~Long}	        \affiliation{\Kent}
\author{J.~Mammei}              \affiliation{\UMASS}  
\author{P.~Markowitz}           \affiliation{\FIU}
\author{T.~Maruyama}            \affiliation{\SLAC}
\author{V.~Maxwell}	        \affiliation{\FIU}
\author{S.~Mayilyan}            \affiliation{\Yerevan}
\author{J.~McDonald}            \affiliation{\SLAC}
\author{R.~Michaels}            \affiliation{\JLab}
\author{K.~Moffeit}             \affiliation{\SLAC}
\author{V.~Nelyubin}            \affiliation{\UVa}
\author{A.~Odian}               \affiliation{\SLAC}
\author{M.~Oriunno}             \affiliation{\SLAC}
\author{R.~Partridge}           \affiliation{\SLAC}
\author{M.~Paolone}             \affiliation{\SCU}
\author{E.~Piasetzky}           \affiliation{\TelAviv} 
\author{I.~Pomerantz}	        \affiliation{\TelAviv}
\author{Y.~Qiang}               \affiliation{\JLab}  
\author{S.~Riordan}             \affiliation{\UMASS} 
\author{Y.~Roblin}              \affiliation{\JLab} 
\author{B.~Sawatzky}            \affiliation{\JLab} 
\author{P.~Schuster}           \thanks{pschuster@perimeterinstitute.ca}  \affiliation{\SLAC}     \affiliation{\Perimeter} 
\author{J.~Segal}               \affiliation{\JLab}
\author{L.~Selvy}	        \affiliation{\Kent}
\author{A.~Shahinyan}           \affiliation{\Yerevan}
\author{R.~Subedi}              \affiliation{\GWU}
\author{V.~Sulkosky}            \affiliation{\MIT}
\author{S.~Stepanyan}           \affiliation{\JLab} 
\author{N.~Toro}                \thanks{ntoro@perimeterinstitute.ca} \affiliation{\Stanford} \affiliation{\Perimeter} 
\author{D.~Walz}               \affiliation{\SLAC}  
\author{B.~Wojtsekhowski}  \thanks{bogdanw@jlab.org} \affiliation{\JLab}
\author{J.~Zhang}               \affiliation{\JLab} 
\date{\today}
\begin{abstract}
We present a search at Jefferson Laboratory for new forces mediated by sub-GeV vector bosons with 
weak coupling $\alpha'$ to electrons. Such a particle $A'$ can be produced in 
electron-nucleus fixed-target scattering and then decay to an $e^+e^-$ pair, producing 
a narrow resonance in the QED trident spectrum. 
Using APEX test run data, we searched in the mass range 175--250~MeV, found no 
evidence for an $A'\to e^+e^-$ reaction, and set an upper limit of $\alpha'/\alpha \simeq 10^{-6}$.  
Our findings demonstrate that fixed-target searches can explore a new, wide, and 
important range of masses and couplings for sub-GeV forces.
\end{abstract}


\maketitle

The strong, weak, and electromagnetic forces are mediated by vector bosons of 
the Standard Model.  
New forces could have escaped detection only if their mediators are 
either heavier than $\mc{O}$(TeV) or quite weakly coupled.  The latter possibility 
can be tested by precision colliding-beam and fixed-target experiments. 
This \emph{letter} presents the results of a search for sub-GeV mediators of weakly coupled
new forces in a test run for the $A'$ Experiment (APEX), which was proposed 
in \cite{Proposal,*Essig:2010xa} based on the general concepts presented in \cite{Bjorken:2009mm}.  

A new abelian gauge boson, $A'$, can acquire a small coupling to
charged particles if it mixes kinetically with the photon \cite{Holdom:1985ag,Galison:1983pa}.  
Indeed, quantum loops of heavy particles with electric and $U(1)'$
charges can generate kinetic mixing and an effective interaction
$\epsilon e A'_\mu J^\mu_{\rm EM}$ of the $A'$ to the 
electromagnetic current $J^\mu_{EM}$,  suppressed relative to the electron charge 
$e$ by $\epsilon \sim 10^{-2}-10^{-6}$ \cite{Essig:2009nc}.  
This mechanism motivates the search for very weakly coupled gauge bosons.
Anomalies related to dark matter~\cite{ArkaniHamed:2008qn,*Pospelov:2008jd}  
and to the anomalous magnetic moment of the muon~\cite{Pospelov:2008zw}
have motivated interest in the possibility of an $A'$ with MeV- to GeV-scale mass.
Gauge bosons in the same mass range arise in several theoretical proposals
 \cite{Fayet:2007ua,Cheung:2009qd,*ArkaniHamed:2008qp,*Morrissey:2009ur}, and their  
couplings to charged matter, $\alpha' \equiv  \epsilon^2 \alpha$
($\alpha= e^2/4\pi$), are 
remarkably weakly constrained \cite{Bjorken:2009mm}.  

\begin{figure}[t]
\begin{center}
\subfigure[]{
\includegraphics [trim = 0mm 5mm 0mm 0mm, width = 0.19\textwidth]{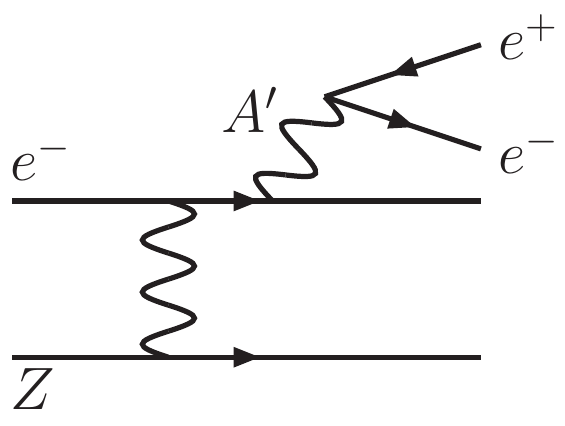}
\label{fig:production.1a}
}\\
\subfigure[]{
\includegraphics [trim = 0mm 5mm 0mm 0mm, width = 0.19\textwidth]{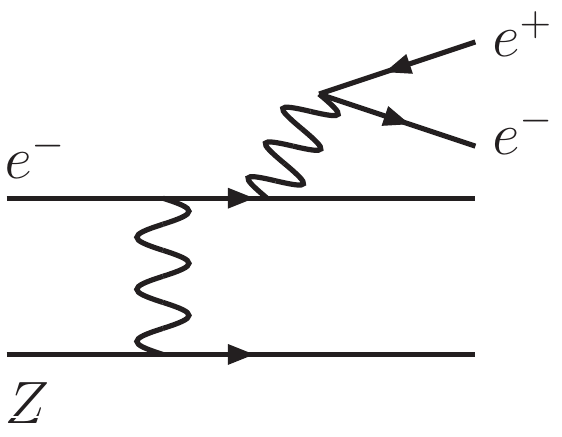}
\label{fig:production.1b}
}
\hskip 10mm
\subfigure[]{
\includegraphics [trim = 0mm 5mm 0mm 0mm, width = 0.20\textwidth]{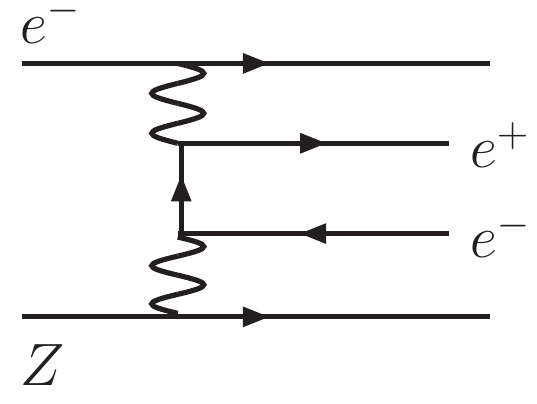}
\label{fig:production.1c}
}
\vspace{-.10 in}
\caption{\emph{Top:} (a) $A'$ production 
from radiation off an incoming $e^-$ beam incident 
on a target consisting of nuclei of atomic number $Z$.  
APEX is sensitive to $A'$ decays to $\epm$ pairs, although decays to 
$\mu^+\mu^-$ pairs are possible for $A'$ masses $m_{A'}>2m_\mu$.  
\emph{Bottom:} QED trident backgrounds: (b) \emph{radiative} tridents and 
(c) \emph{Bethe-Heitler} tridents.
}
\label{fig:production}
\end{center}
\vspace{-.30 in}
\end{figure}

The simplest scenario, in which the $A'$ decays directly to ordinary matter, 
can be tested in electron and proton fixed-target experiments 
\cite{Bjorken:2009mm,Freytsis:2009bh,Batell:2009di,*Essig:2010gu} and at $e^+e^-$ and hadron colliders  
\cite{Essig:2009nc,Cheung:2009qd,Reece:2009un,*:2009cp,Batell:2009yf,*Aubert:2009pw,*Wojtsekhowski:2009vz,*Abazov:2009hn,*Abazov:2010uc,Archilli:2011nh}.  
Hidden sector collider phenomenology has also been explored in detail in e.g.~\cite{Strassler:2006im}. 
Electron fixed-target experiments are uniquely suited to probing the sub-GeV mass 
range because of their high luminosity, large $A'$ production cross section, and favorable kinematics.   
Electrons scattering off target nuclei can radiate an $A'$, which 
then decays to $e^+e^-$, see Fig.~\ref{fig:production}.  
The $A'$ would then appear as a 
narrow resonance in the $\epm$ invariant 
mass spectrum, over the large background from quantum electrodynamics (QED) trident processes.
APEX is optimized to search for such a resonance using 
Jefferson Laboratory's Continuous Electron Beam Accelerator
Facility and two High Resolution Spectrometers (HRSs) in Hall A~\cite{Alcorn:2004sb}.  

The full APEX experiment proposes to probe couplings $\alpha'/\alpha \gtrsim 10^{-7}$ and masses 
$m_{A'}\sim50-550~\MeV$, a considerable improvement in cross section sensitivity over previous experiments
in a theoretically interesting region of parameter space. 
Other electron fixed-target experiments are planned at Jefferson Laboratory, 
including the Heavy Photon Search (HPS)~\cite{HPS} and DarkLight
\cite{Freytsis:2009bh} experiments; 
at MAMI \cite{Merkel:2011ze}; and at DESY (the HIdden Photon Search (HIPS) \cite{Andreas:2010tp,*HIPS}).  

\begin{figure}[t]
\begin{center}
\includegraphics [trim = 0mm 0mm 0mm 0mm, width = 0.48\textwidth]{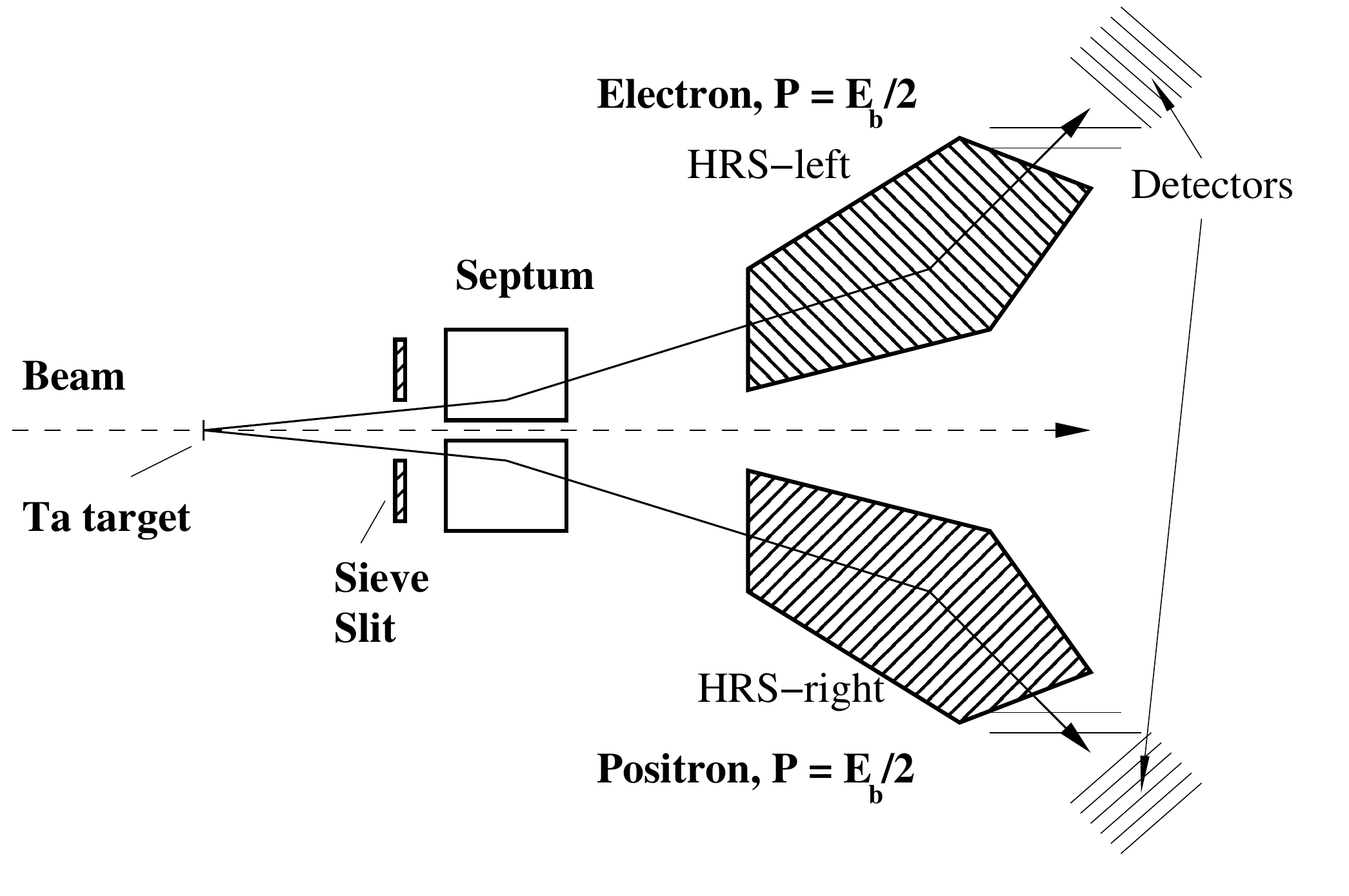}
\caption{The layout of the APEX test run.  An electron beam (left-to-right) is incident on a 
thin tantalum foil target. 
Two septum magnets of opposite polarity deflect charged particles to larger angles into 
two vertical-bend high resolution spectrometers (HRS) set up to select electrons and positrons,  each 
carrying close to half the incoming beam energy.
The HRSs contain detectors to accurately measure the momentum, direction, and identity of the particles. 
Insertable sieve slit plates located in front of the septum magnets were
used for calibration of the spectrometer magnetic optics.
}
\label{fig:layout}
\end{center}
\vspace{-.25 in}
\end{figure}

We present here the results of a test run for APEX that took place at Jefferson Laboratory in July 2010.  
The layout of the experiment is shown in Fig.~\ref{fig:layout}. 
The distinctive kinematics of $A'$ production motivates the choice of configuration.  
The $A'$ carries a large fraction of the incident beam energy, $E_{\rm b}$, is produced at angles 
 $\sim (m_{A'}/E_{\rm b})^{3/2} \ll 1$, and decays to an $\epm$ pair with a typical angle of 
$m_{A'}/E_{\rm b}$.
A symmetric configuration with the $e^-$ and $e^+$ each carrying nearly 
half the beam energy mitigates QED background while maintaining high signal efficiency.

The test run used a $2.260 \pm 0.002$~GeV  electron beam with an intensity up 
to 150~$\mu$A incident on a tantalum foil of thickness 22~mg/cm$^2$.
The HRSs' central momenta were $\simeq$1.131~GeV with a momentum acceptance of $\pm4.5$\%.
Dipole septum magnets between the target and the HRS aperture allow the detection of 
$e^-$ and $e^+$ at angles of 5$\degr$ relative to the incident beam. 
Collimators present during the test run reduced the solid angle acceptance of each spectrometer
from a nominal 4.3~msr to $\simeq 2.8~(2.9)$ msr for the left (right) HRS.

The two spectrometers are equipped with similar detector packages.
Two vertical drift chambers, each with two orthogonal tracking planes, provide 
reconstruction of particle trajectories. 
A segmented timing hodoscope and a gas Cherenkov counter (for  $e^+$ identification)
are used in the trigger.  A two-layer lead glass calorimeter provides further offline particle identification.   
A single-paddle scintillator counter is used for timing alignment. 

Data were collected with several triggers:  
the single-arm triggers produced by the hodoscope in either arm, a double
coincidence trigger produced by a 40-ns wide overlap between the hodoscope signals 
from the two arms, and a triple coincidence trigger consisting of the double coincidence signal and
a gas Cherenkov signal in the positron (right) arm. 
Single-arm trigger event samples are used for optics and acceptance calibration, described below.
The double coincidence event sample, which is dominated by accidental $e^-\pi^+$
coincidences, is used to check the angular and momentum acceptance of the spectrometers.  
These  $e^-\pi^+$ coincidences are largely rejected in the triple
coincidence event sample by the requirement of a gas Cherenkov signal in the positron arm.    

The reconstruction of $e^+$ and $e^-$ trajectories at the target was calibrated
using the sieve slit method, see~\cite{Offermann1987298,  Alcorn:2004sb}. 
The sieve slits --- removable tungsten plates with a grid of holes drilled through at known
 positions --- are inserted between the target and the septum magnet
 during the calibration runs.
 In this configuration, data were taken with a 1.131~GeV and a 2.262~GeV incident electron beam.  
Using the reconstructed track positions and angles as measured in the vertical drift chambers, and 
the spectrometer's optical transfer matrix, the positions at the sieve slit were calculated.
The parameters of the optical transfer matrix are then optimized 
to produce the best possible overlap with the sieve holes positions, and this
corrected matrix is applied to event reconstruction.
Only events within calibrated acceptance are used in the final analysis.  

The final event sample is selected from the coincidence sample defined above by imposing 
a 12.5-ns time window between the electron arm trigger and the positron
arm gas Cherenkov signals (no off-line corrections were applied), 
requiring good quality tracks in the vertical drift chambers of both arms, 
and the acceptance selection described above.
Lastly, we demand that the sum of $e^+$ and $e^-$ energies not exceed the beam-energy threshold 
for true coincidence events of 2.261 GeV, which reduces accidental coincidences. 
This final sample of 770,500 events consists almost entirely of true $e^+e^-$ coincidence 
events with only 0.9\% contamination by meson backgrounds, 
and 7.4\% accidental $\epm$ coincidence events.  

\begin{figure}[t]
\begin{center}
\includegraphics [trim = 0mm 5mm 0mm 10mm, width = 0.48\textwidth]{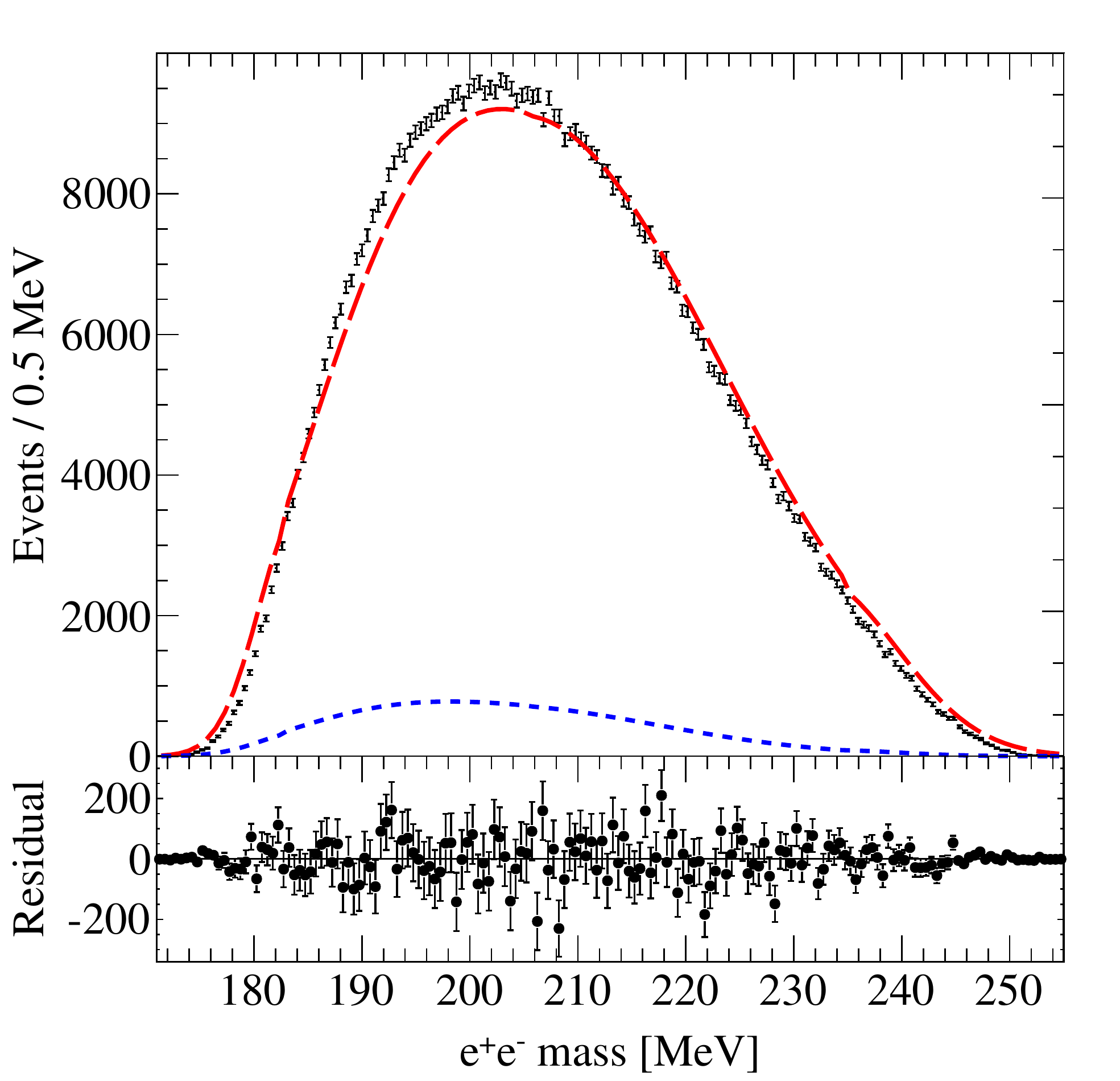}
\caption{\emph{Upper panel:} The invariant mass spectrum of $\epm$ pair events in the final 
event sample (black points, with error bars), accidental $\epm$ coincidence events (blue short-dash line), 
and the QED calculation of the trident background added to the
accidental event sample (red long-dash line).
\emph{Lower panel:} the bin-by-bin residuals with respect to a 10-parameter fit to the 
global distribution (for illustration only, not used in the analysis).
}
\label{fig:mass}
\end{center}
\vspace{-.30 in}
\end{figure}

The experimental data were compared with a calculation of the 
leading order QED trident process using MadGraph and MadEvent~\cite{Alwall:2007st}. 
MadEvent was modified to account for nucleus-electron kinematics and to use
the nuclear elastic and inelastic form factors in \cite{Kim:1973he}. 
The invariant mass spectrum of the calculated coincident event sample 
overall normalized to the data is shown in Fig.~\ref{fig:mass}. 
Overall trident rates from our calculations for the test run configuration, 
accounting for acceptance, agree within a few percent with data.  
Likewise, the differential momentum and angular distributions agree within
$5-10\%$. 
The remaining discrepancies are consistent with uncertainties in
the multi-dimentional momentum-angular acceptance and 
detector efficiency effects not included in our comparison. 

The sensitivity to $A'$ depends critically on precise reconstruction of the invariant mass of $\epm$ pairs.
Due to the excellent HRS relative momentum resolution of $O(10^{-4})$,
the mass resolution is dominated by three contributions to the angular resolution:  
scattering of the $\epm$ inside the target, track measurement errors by 
the HRS detectors, and imperfections in the magnetic optics reconstruction matrix.  
Multiple scattering in the target contributes 0.37~mrad to the vertical and 
horizontal angular resolutions for each particle.
Track measurement uncertainties contribute typically 0.33~(1.85)~mrad to the 
horizontal (vertical) angular resolution in the left HRS and 0.43~(1.77) in the right HRS. 
Magnetic optics imperfections in both HRSs were found to contribute typically 
0.10~(0.22)~mrad to the horizontal (vertical) angular resolution.  
Because calibration of the magnetic optics was performed using only $e^-$, and not $e^+$, 
there is a possibility of additional aberrations in the positron arm.
An upper limit for possible aberrations of 0.5~mrad was obtained from angular correlations 
in $H(e,e'p)$ experiments with the HRS and the calculations of the septum magnetic field.
Accounting for these effects, we determine the combined mass resolution (rms) to be between
$0.85$ and $1.11 \MeV$, depending on the invariant mass.  
Finally, the uncertainty in the absolute angle between the two sieve slits introduces 
a 1\% uncertainty in the absolute mass scale but does not affect the mass resolution.

The starting point for the $A'\rightarrow \epm$ search is the invariant mass distribution 
of the coincident event sample, shown in black in Fig.~\ref{fig:mass}.  
Also shown is the accidental $\epm$ coincidence event sample in blue, and 
the QED calculation of the trident background added to the accidental sample in red.
For illustration, we show the bin-by-bin residuals with respect to a 10-parameter fit to the global 
distribution, although we do not use this in the analysis. 
The analysis code, described below, was tested and optimized on our
simulated data and on a $10\%$ sample of the experimental data to avoid possible bias. 

\begin{figure}[t]
\begin{center}
\includegraphics [trim = 23mm 12mm 30mm 7mm, width = 0.35\textwidth]{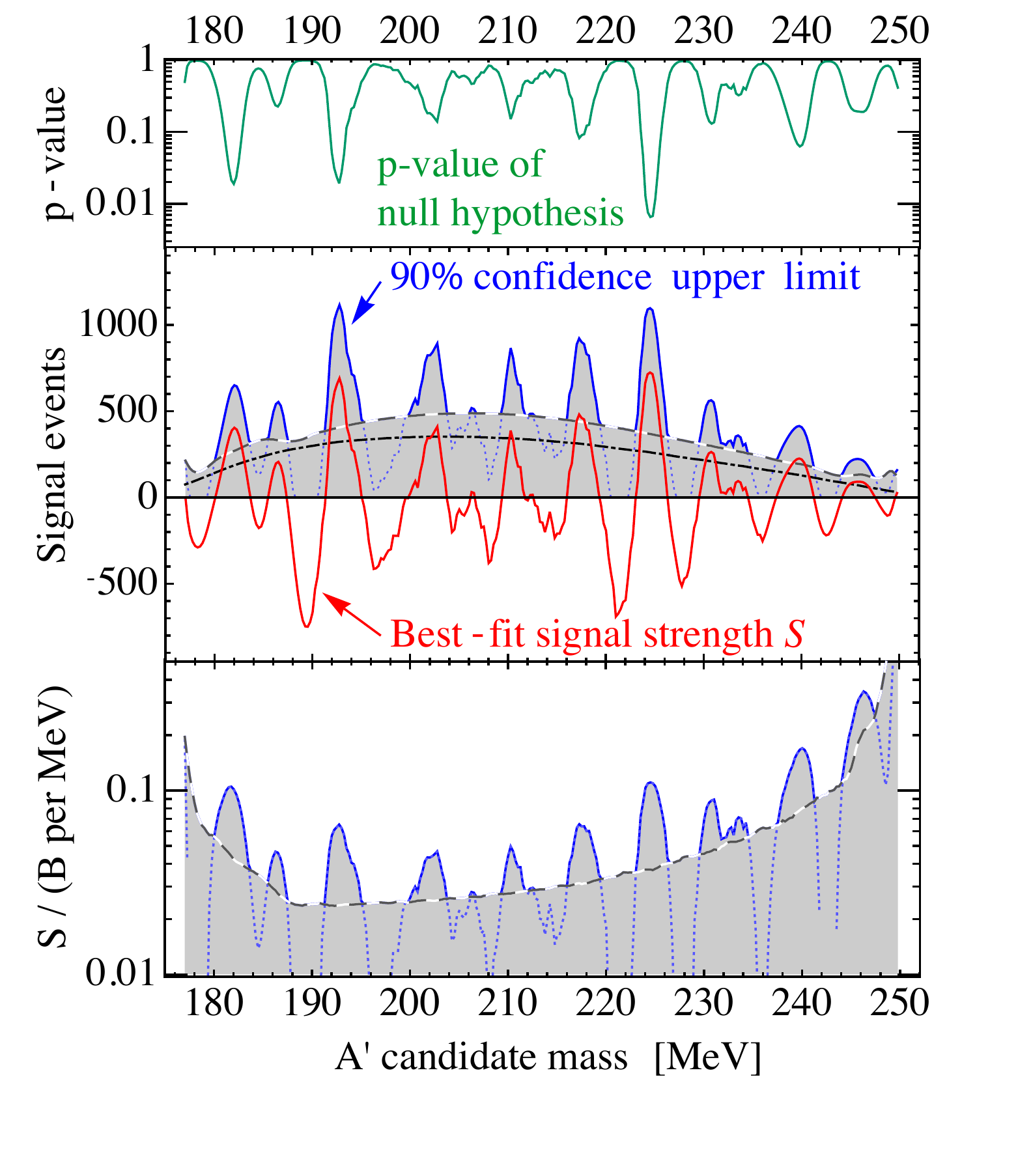}
\caption{
\emph{Top:} Background-only model $p$-value versus $A'$ mass. 
\emph{Middle:}  Shaded gray region denotes 90\% confidence limit, 
50\% power-constrained allowed region~\cite{Cowan:2011an}.  
90\% confidence upper limit is shown in solid blue (dotted blue) when 
it is above (below) the expected limit (gray dashed). 
Red solid line denotes the best-fit for the number of signal events $S$.  
For comparison, dot-dashed line indicates contribution of statistical 
uncertainty to expected sensitivity, if background shape were known exactly.  
\emph{Bottom:} 90\% confidence,  50\% power-constrained, and expected limits
as above, here quoted in terms of ratio of signal strength upper-limit
to the QED background, B, in a 1-MeV window around each $A'$ mass hypothesis.
}
\label{fig:result1}
\end{center}
\vspace{-.30 in}
\end{figure}

We found that a linear sideband analysis is not tenable in light of
the high statistical sensitivity of the experiment and
the appreciable curvature of the invariant mass distribution; it 
suffers from $O(1)$ systematic pulls, which can
produce false positive signals or overstated sensitivity.
Instead, a polynomial background model plus a Gaussian signal of 
$S$ events (with mass-dependent width corresponding to the mass 
resolution presented above) 
is fit to a window bracketing each candidate $A'$ mass. 
The uncertainty in the polynomial coefficients incorporates the
systematic uncertainty in the shape of the background model.   
Based on extensive simulated-experiment studies, a 7th-order polynomial
fit over a 30.5 MeV window was found to achieve near-minimum uncertainty 
while maintaining a potential bias below 0.1 standard deviations across the mass spectrum.  
A symmetric window is used, 
except for candidate masses within 15 MeV of the upper or lower boundaries, 
for which a window of equal size touching the boundary is used.  
A binned profile likelihood ratio (PLR) is computed as a function of signal strength 
$S$ at the candidate mass, using 0.05 MeV bins. 
The PLR is used to derive the local probability ($p$-value) at $S=0$ (i.e. the probability
of a larger PLR arising from statistical fluctuations in the background-only model) 
and a 90\%-confidence upper limit on the signal.  
We define the sensitivity of the search in terms of a 50\% 
power-constraint~\cite{Cowan:2011an}, 
which means we do not regard a value of $S$ as excluded if it 
falls below the expected limit. 
This procedure is repeated in steps of $0.25 \MeV$.   
A \emph{global} $p$-value, corrected for the ``look-elsewhere effect'', 
(the fact that an excess of events \emph{anywhere} in the range can 
mimic a signal), is derived from the lowest local $p$-value observed over the
full mass range, and calibrated using simulated experiments.  

We find no evidence of an $A'$ signal. 
The $p$-value for the background model and upper bound on the absolute yield of 
$A'\rightarrow \epm$ signal events (consistent with the data and 
background model) are shown in Fig.~\ref{fig:result1}.
The invariant-mass-dependent limit is $\simeq 200-1000$ signal events at 90\% confidence.  
The most significant excess, at $224.5 \MeV$, has a local $p$-value of 0.6\%; the 
associated global $p$-value is 40\%  
(i.e. in the absence of a signal, 40\% of prepared experiments 
would observe a more significant effect due to fluctuations).  

\begin{figure}[t]
\begin{center}
\includegraphics [trim = 0mm 10mm 5mm 10mm, width = 0.45\textwidth]{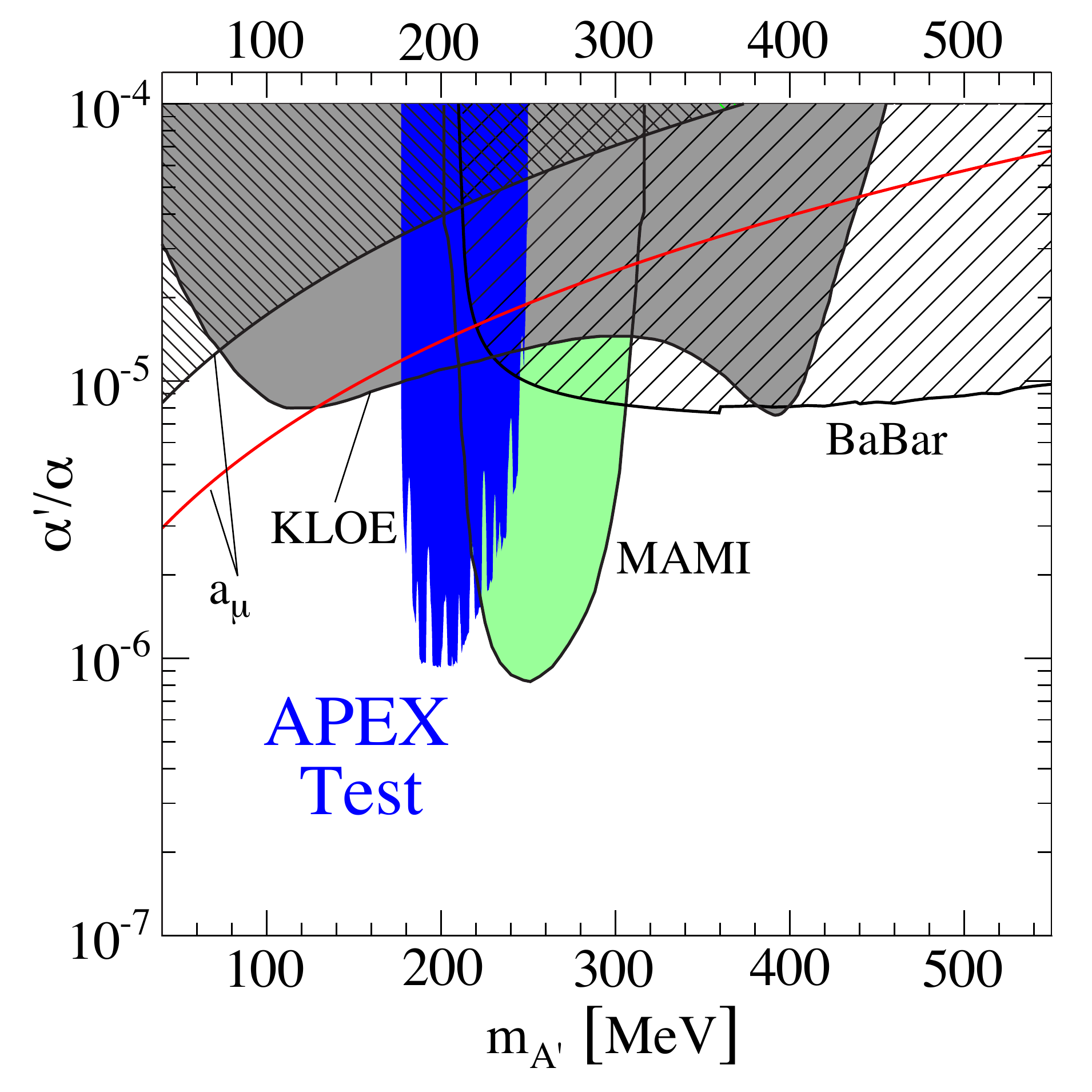}
\caption{
The 90\% confidence upper limit on $\alpha ' / \alpha$ 
versus $A'$ mass for the APEX test run (solid blue). 
Shown are existing 90\% confidence level limits from the muon anomalous magnetic moment $a_\mu$ 
(fine hatched)~\cite{Pospelov:2008zw},  
KLOE (solid gray)~\cite{Archilli:2011nh}, 
the result reported by Mainz (solid green)~\cite{Merkel:2011ze}, and
an estimate using a BaBar result (wide hatched)~\cite{Bjorken:2009mm,Reece:2009un,*:2009cp}.
Between the red line and 
fine hatched region, the $A'$ can explain the observed discrepancy between the
calculated and measured muon anomalous magnetic moment \cite{Pospelov:2008zw} at 90\% confidence 
level.
The full APEX experiment will roughly cover the
entire area of the plot.}
\label{fig:result2}
\end{center}
\vspace{-.25 in}
\end{figure}

To translate the limit on signal events into an upper limit on the coupling $\alpha'$ 
with minimal systematic errors from acceptance and trigger efficiencies, 
we use a ratio method, normalizing $A'$ production to the measured QED trident rate.  
We distinguish between three components of the QED trident background: \emph{radiative} tridents 
Fig.~\ref{fig:production} (b), \emph{Bethe-Heitler} tridents Fig.~\ref{fig:production} (c), 
and their interference diagrams (not shown).
The $A'$ signal and \emph{radiative} trident fully differential cross sections are simply related 
\cite{Bjorken:2009mm}, and the ratio $f$ of the radiative-only cross section to the 
full trident cross section can be reliably computed in Monte Carlo: 
$f$ varies linearly from 0.21 to 0.25 across the APEX mass range, with a systematic uncertainty of 
0.01, which dominates over Monte Carlo statistics and possible next-to-leading order QED effects.
The 50\% power-constrained limit on signal yield $S_{max}$ and trident
background yield per unit mass, $\Delta B/\Delta m$, evaluated in a 1 MeV range around $m_{A'}$,
determines an upper limit on $\alpha'/\alpha$,
\begin{equation*}
\left( \frac{\alpha'}{\alpha} \right)_{max}= 
\left( \frac{ S_{max}\,/\,m_{A'}}{f \, \cdot \Delta B/\Delta m} 
\right) \times \left( \frac{2\, N_{\text{eff}}\, \alpha}{3 \,\pi}\right) ,
\end{equation*}

where $N_{\rm eff}$ counts the number of available decay channels 
($N_{\rm eff} =  1$ for $m_{A'}<2m_{\mu}$, and increases to $\simeq 1.6$ at 
$m_{A'}\simeq 250$ MeV).
The resulting limit, accounting in addition for contamination of
the background by accidentals, is shown in Fig.~\ref{fig:result2}.  

In summary, the APEX test run data showed no significant signal
of $A'\to e^+e^-$ electro-production in the mass range 175--250 $\MeV$. 
We established an upper limit of $\alpha'/\alpha \simeq 10^{-6}$ at 90\% confidence.  
All aspects of the full APEX experiment outlined in \cite{Proposal} have 
been demonstrated to work. 
The full experiment plans to run at several beam energies,  
have enhanced mass coverage from a 50-cm long multi-foil target, 
and acquire $\sim 200$ times more data than this test run,
extending our knowledge of sub-GeV force. 

\begin{acknowledgments}
The APEX collaboration thanks the JLab technical staff for their  
tremendous support during the brief test run.
This work was supported by the U.S. Department of Energy. 
Jefferson Science Associates, LLC, operates Jefferson Lab for 
the U.S. DOE under U.S. DOE contract DE-AC05-060R23177.
This work was also supported in part by the U.S. Department of Energy under
contract number DE-AC02-76SF00515 and by the National Science 
Foundation under Grant No. NSF PHY05-51164.  
\end{acknowledgments}

\bibliography{apex_arXiv_v2}
\end{document}